\documentclass[prc,preprint,aps]{revtex4}
\usepackage{graphicx}
\newcommand{\be}{\begin{equation}}
\newcommand{\ee}{\end{equation}}
\newcommand{\ben}{\begin{eqnarray}}
\newcommand{\een}{\end{eqnarray}}

\begin{document}
\draft
\title{On the Cut-Off Prescriptions Associated with Power-Law Generalized 
Thermostatistics}

\author{A.M. Teweldeberhan$^{1}$,  A.R. Plastino $^{1,\,2,\,3}$ and H.G. 
Miller$^1$}

\affiliation{$^1$Department of Physics, University of Pretoria \\
Pretoria 0002,South Africa}

\affiliation{$^2$Departament de F\'{\i}sica, Universitat de les Illes Balears, 
07071 Palma de Mallorca, Spain}

\affiliation{$^3$Facultad de Ciencias Astronmicas y Geof\'{\i}sicas, UNLP and CONICET\\
C.C. 727, 1900 La PLata, Argentina}




\date{\today}

\begin{abstract}

 We revisit the cut-off prescriptions which are
needed in order to specify completely the form of Tsallis' maximum entropy 
distributions. For values of the Tsallis entropic parameter $q>1$
we advance an alternative cut-off prescription and discuss some of its basic 
mathematical properties. As an illustration of the new cut-off prescription we
consider in some detail the $q$-generalized quantum distributions which have 
recently been shown to reproduce various experimental results related to high 
$T_c$ superconductors.
  \end{abstract}  


\maketitle



\section{Introduction}


There is a growing body of evidence 
\cite{LSS00,LK98,PP95,B98,FD99,CLPT97,
T00,MTL00,TTL02,BR02a,BR02b,HD94,
B95,B00,B02a,B02b,B04,AA01,B01,ATM02,UMK01,
BRAZIL,AO01,tsallis1,tsallispage} 
indicating that there are 
important systems and processes in physics, 
biology, economics, and other fields, which are described by statistical 
distributions of the Tsallis' maximum entropy form. These distributions are 
obtained from the extremalization of Tsallis' entropic functional under the 
constraints imposed by normalization
and the mean values of a (in general small) set of relevant quantities which 
are regarded as input information. The Tsallis entropic measure is given by 
\cite{T88}

\begin{equation}
S_q \, = \, \frac{k}{q-1} \, \left( 1 -  \sum^w_{i=1} p^q_i \right)
\end{equation}

\noindent
 where $k$ is a positive constant (from here on set equal to 1), $w$ is
the total number of microstates in the system, $\{p_i, \,\, i=1,\ldots, w\}$
are the associated probabilities, and the Tsallis parameter $q$ is any real
number. It is straight forward to verify that the usual Boltzmann-Gibbs (BG)
logarithmic entropy,  $S = - \sum_i  p_i \ln p_i$, is recovered in the limit
$q\rightarrow 1$.  Tsallis' maximum entropy distributions have been found to be 
relevant, for instance, in various astrophysical scenarios \cite{LSS00,LK98}
and in the study of the phenomenon of non-linear diffusion 
\cite{PP95,B98,FD99}. The description of the behaviour of chaotic
maps at the threshold of chaos constitutes another successful field
of application for the Tsallis formalism 
\cite{CLPT97,T00,MTL00,TTL02,BR02a,BR02b}. It is
important to stress that some of the aforementioned developments
involve a quantitative agreement between experimental data and
theoretical models based on Tsallis maximum entropy distributions. For
instance, it was experimentally found that pure electron plasmas
in Penning traps relax to metastable states whose radial density
profiles do not maximize the Boltzmann-Gibbs entropy \cite{HD94}. However,
Boghosian showed that the observed profiles are well described by
a Tsallis distribution with $q$ close to $1/2$ \cite{B95}.
Beck's recent investigations on fully developed turbulent flows
constitute another interesting application \cite{B00,B02a,B02b,B04}.
General overviews on Tsallis' formalism and its diverse applications
can be found in \cite{BRAZIL,AO01,tsallis1}. An updated bibliography 
is available in \cite{tsallispage}.

The formal solutions to Tsallis' maximum entropy
variational problem, the so called $q$-maxent distributions, are not always 
positive real
numbers. In order to guarantee the real and positive character of these 
$q$-maxent distributions it is necessary to introduce appropriate cut-off 
prescriptions. The aim of the present work is
to discuss one possible cut-off prescription for the case of Tsallis parameter 
$q>1$. Recent studies by various authors suggest that there 
are two classes of non-extensive descriptions, corresponding, respectively, 
to $q<1$ and $q>1$. For instance, earlier applications to low-dimensional
dissipative maps at the edge of chaos 
yield $q$-values less then $1$ \cite{CLPT97},
while it was latter realized that there is another class of $q$-values 
greater than $1$ \cite{MTL00}. 
A similar situation occured in the case of turbulence \cite{B04,AA01}.
 So far the $q<1$ case has been studied in greater detail, and is 
consequently better understood, than the $q>1$ case. We believe that 
the present work may contribute to a deeper understanding of 
this latter case.

The paper is organized as follows. In section II
we review the Tsallis cut-off prescription.
In section III we propose an alternative prescription for the case $q>1$. In 
section IV
we illustrate our prescription with a $q$-generalized Fermi-Dirac distribution 
and
we prove that, in this case, the present prescription leads to a 
thermodynamically consistent formalism. Some conclusions are
drawn in section V.

\section{Tsallis' Cut-off Prescription}

 When, in addition to normalization, one only has the
mean energy constraint, Tsallis' distribution can be
parameterized as \cite{CT91}

\be \label{tsadis}
p_i \, = \, \frac{1}{Z_q}
\left[
1 \, - \, (1-q)\beta \epsilon_i
\right]^{\frac{1}{1-q}},
\ee 

\noindent
where $q$ is a real number (Tsallis' parameter), $\epsilon_i$ is the energy of 
microstate $i$, and the partition function $Z_q$ is an appropriate 
normalization constant. This distribution can be regarded as the 
$q$-generalization of the
Gibbs canonical distribution. In order
to guarantee that the microstates probabilities $p_i$ are non-negative real 
numbers, it is necessary to supplement expression
(\ref{tsadis}) with an appropriate prescription for treating
negative values of the quantity under square brackets. That is,
we need a prescription for the value of $p_i$ when

\be \label{inequ}
1 \, - \, (1-q)\beta \epsilon_i \, < \, 0.
\ee

\noindent
The simplest possible prescription, and the one usually
adopted, is to set $p_i=0$ whenever inequality (\ref{inequ}) holds 
\cite{T88,CT91}. This rule,
 usually referred to as ``Tsallis' cut-off prescription", may seem at first sight 
just an ad-hoc solution to the above ``negativity issue". However, in many 
specific scenarios, Tsallis'
cut-off prescription turns out to be a physically sensible one. It is 
instructive to mention a few examples.
In the polytropic models of self-gravitating $N$-body systems, which are 
described by Tsallis' maximum entropy distributions, the cut-off corresponds to 
the escape
velocity from the system \cite{PP93a}. Other interesting examples involve the 
so-called $q$-gaussian distributions,

\be \label{qgauss}
\frac{1}{Z_q}
\left[
1 \, - \, (1-q)\beta x^2
\right]^{\frac{1}{1-q}},
\ee

\noindent
which reduce to the standard gaussian distribution in the limit $q\rightarrow 
1$. For $q<1$ there are important
non-linear diffusion and Fokker-Planck equations, with multiple applications in 
diverse fields, which admit  exact analytical solutions of the $q$-gaussian 
form {\it with the Tsallis cut-off } \cite{PP95,B98,FD99}. Finally, it is 
possible to construct one dimensional quantum mechanical potential functions, 
exhibiting interesting properties related to shape invariance and, again, 
admitting exact ground state wave functions of the $q$-gaussian form {\it with 
the Tsallis cut-off} \cite{RPGCP00}. It is clear from the above examples that 
Tsallis' cut-off prescription is indeed a physically reasonable one. This 
prescription may not constitute, however, the complete and final answer to the 
negativity issue. In order to clarify this, we have first to realize that 
Tsallis' prescription covers two very different situations:

\begin{itemize} \item{(a) When $q<1$ there is a special (positive) value of the 
quantity $\beta \epsilon$, $(\beta \epsilon)_c$,
for which the probability distribution becomes zero. In this case the 
probability distribution is set equal to zero
for $\beta \epsilon > (\beta \epsilon)_c$. With this prescription the 
probability distribution remains a continuous function of $\beta \epsilon$. 
Even the first derivative of $p$ with respect to $\beta \epsilon$ is, for a 
certain range of $q$-values, continuous at the cut-off point.}
\item{(b) When $q>1$ a completely different picture obtains. Now there is a 
particular (negative) value of
$\beta \epsilon$,  $(\beta \epsilon)_c$, such that
when $\beta \epsilon$ approaches $(\beta \epsilon)_c$
from the left $p\longrightarrow +\infty$. In this case
Tsallis' cut-off rule prescribes that $p$ is to be set equal to zero for all 
$\beta \epsilon < (\beta \epsilon)_c$.
}
\end{itemize}

Two comments are in order. First of all, most of the
concrete physical realizations of Tsallis' cut-off
that have so far been studied (and, in particular, the three examples 
previously mentioned by us)
correspond to the $q<1$ (case (a)) instance. Secondly, the cut-off prescription 
is much less palatable in case (b) than in case (a). In the 
latter case, 
$p(\beta \epsilon)$
is a continuous function while in the former case
it jumps from $+\infty$ to $0$ at the cut-off point.
The main aim of our present contribution is to discuss a possible alternative 
to the $q>1$ case
of the cut-off prescription.

\section{Alternative Cut-off Prescription for $q>1$.}

 Tsallis' maximum entropy distributions can be conveniently written in terms of 
the $q$-generalized exponential function

\begin{equation}
{\rm e}_q(x)=\left\{ 
\begin{array}{ll}
\left[1+(1-q)x\right]^{\frac{1}{1-q}}\,, &\quad \left[1+(1-q)x \right] >0 \vspace{0.5cm} \\
0\,, & \quad \left[1+(1-q)x \right] \le 0.
\end{array}
\right.
\end{equation}
Notice that this last expression contains Tsallis' cut-off condition, which is
absorbed into the definition of the $q$-exponential function ${\rm e}_q(x)$.
In terms of the $q$-generalized exponential, Tsallis' distribution is 

\be
p_i \, = \, {\rm e}_q(-\beta \epsilon_i).
\ee

We are going to introduce now an alternative generalization ${\rm \tilde e}_q(x)$ of 
the exponential function, defined in the following way. For $q<1$ we set 
${\rm \tilde e}_q(x)= {\rm e}_q(x)$. And for $q>1$, we propose

\begin{equation} \label{exqu}
{\rm \tilde e}_q(x)=\left\{\begin{array}{ll}
\left[1+(q-1)x\right]^{\frac{1}{q-1}}\,, &\quad x>0 \vspace{0.5cm} \\
\left[1+(1-q)x\right]^{\frac{1}{1-q}}\,, &\quad x \le 0
\end{array} \right.
\end{equation}

\noindent
The function ${\rm \tilde e}_q(x)$ has, for $q>1$, some desirable properties. First 
of all, ${\rm \tilde e}_q(x)$ complies with the ``exponential-like" relation

\begin{equation}
{\rm \tilde e}_q(x)\,.\,{\rm \tilde e}_q(-x)=1.
\end{equation} 

\noindent
On the other hand, ${\rm \tilde e}_q(x)$ is clearly continuous at $x=0$ (See Figure 
1). Furthermore,
we have 

\[
\frac{d}{dx}\left([1+(q-1)x]^{\frac{1}{q-1}}\right)=[1+(q-1)x]^{\frac{2-q}{q-1}}\quad \longrightarrow 1\quad {\rm as}\quad x \rightarrow 0^{+},
\]

\noindent
and

\begin{equation}
\frac{d}{dx}\left([1+(1-q)x]^{\frac{1}{1-q}}\right)=[1+(1-q)x]^{\frac{q}{1-q}}\quad \longrightarrow 1\quad {\rm as}\quad x \rightarrow 0^{-},
\end{equation}

\noindent
implying that $\frac{d}{dx}{\rm \tilde e}_q(x)$ is, for $q>1$, continuous at $x=0$. 

Now, the alternative cut-off prescription that we want to consider is 
tantamount to adopting  for the $q$-generalized Gibbs ensemble a distribution 
of the form,

\be
p_i \, = \, {\rm \tilde e}_q(-\beta \epsilon_i).
\ee

\noindent
The generalized exponential function ${\rm \tilde e}_q(x)$ 
allows them to express our cut-off prescription, and the 
corresponding generaliztion of the canonical distribution
in a compact form. Our aim in the present {\it Letter}
is to explore some important {\it physical} consequences 
of our cut-off prescription. In particular, we want to address 
its thermodynamic consistency. It would be interesting to 
investigate in detail the mathematical properties of 
the function ${\rm \tilde e}_q(x)$ but, of course, 
that is not our aim here. For our present purposes 
we only need the basic features of
${\rm \tilde e}_q(x)$ already mentioned. A similar situation
arose in the case of Tsallis' generalized exponential
$e_q(x)$: this function was introduced as a compact
and elegant notation for Tsallis' maximum entropy distributions.
The extensive literature on the purely mathematical properties of
$e_q(x)$ only appeared afterwards.

\section{The Generalized Fermi-Dirac Distribution.}

 \subsection{The Standard Maximum Entropy Principle for Quantum Distributions}

The quantum mechanical distributions can be obtained from a maximum entropy
principle based on the entropic measure (the upper signs corresponding to
bosons and lower one to fermions) \cite{KK,S,PPMU04}

\begin{equation} \label{standard1}
S = -\sum_i \Bigl[\bar{n}_i\ln \bar{n}_i \mp (1\pm \bar{n}_i)\ln (1 \pm
\bar{n}_i)\Bigr],
\end{equation}

\noindent where $\bar{n}_i$ denotes the number of particles in the i$^{th}$ energy level with energy $\epsilon_i$. The extremalization of the above measure under the constraints imposed by the total number of particles,

\begin{equation} \label{standard2}
\sum_i\bar{n}_i \, = \, N,
\end{equation}

\noindent
 and the total energy of the system,

\begin{equation} \label{standard3}
\sum_i\bar{n}_i \epsilon_i \, = \, E,
\end{equation}

\noindent leads to the standard quantum distributions,

\begin{equation} \label{standard4}
\bar{n}_i=\frac{1}{ \exp{\beta(\epsilon_i-\mu)} \mp 1}.
\end{equation}

\noindent In the above equation the minus
sign corresponds to the Bose-Einstein
distribution and the plus sign corresponds to
the Fermi-Dirac one.

 \subsection{The Nonextensive Maximum Entropy Principle for Fermions}

 In order to deal with non extensive scenarios (characterized by $q\ne 1 $)
we propose the extended measure of entropy,

\begin{equation} \label{qentrop2}
S_q^{(F)}[\bar{n}]=\sum_i \left[\left(\frac {\bar{n}_i-\bar{n}_i^q} 
{q-1}\right)+\left(\frac{(1-\bar{n}_i)-(1-\bar{n}_i)^q} {q-1}\right)\right].
\end{equation}

\noindent
That the entropic functional (\ref{qentrop2}) constitutes a
natural generalization of (\ref{standard1}) can be easily
realized if we express it in terms of the so-called ``$q$-logarithms",

\be \label{qloga1}
 S_q^{(F)}[\bar{n}] = - \sum_i 
\left[\
\bar{n}_i^q {\rm ln}_q(\bar{n}) + \left(1- \bar{n}_i \right)^q
{\rm ln}_q \right(1 -\bar{n}_i\left)
\right],
\ee

\noindent
where the $q$-logarithm is defined as \cite{AO01,tsallis1,tsallispage}

\be \label{qloga2}
{\rm ln}_q(x) \, = \, (1-q)^{-1} \left(x^{1-q} - 1\right), \,\,\,\,\,\,\,\,\,\, (x>1).
\ee

\noindent In the limit $q\rightarrow 1$ (\ref{qloga1})
reduces to (\ref{standard1}). However, the main 
physical motivation for introducing the measure 
(\ref{qentrop2}) is that it leads, via the maximum 
entropy principle, to quantum distribution functions
that have been very successful in the study of
a concrete and important physical phenomena: high
$T_C$ superconductivity \cite{UMK01}. Furthermore, 
as we shall presently explain, the formulation 
of a variational principle in terms of (\ref{qentrop2})
allows to prove the thermodynamical consistency of
the cut-off prescription used in \cite{UMK01}.

The relevant constraints leading to the $q$-generalized
quantum distributions are the total number of particles,

\begin{equation} \label{numeroq}
\sum_i \bar{n}_i^q=N, \end{equation}

\noindent
and the total energy,

\begin{equation} \label{energiq}
\sum_i \bar {n}_i^q \,\epsilon_i=E. \end{equation}

\noindent
The extremalization of the entropic measure (\ref{qentrop2})
under the constraints (\ref{numeroq}) and (\ref{energiq}) leads to the 
variational problem  

\begin{equation} \label{variatiq}
\delta\left\{S_q^{(F)}[\bar{n}]+\alpha\left(N-\sum_i 
\bar{n}_i^q\right)+\beta\left(E-\sum_i \epsilon_i\,\bar{n}_i^q\right)\right\}=0,
\end{equation}

\noindent
whose formal solution is

\ben \label{formalsoq}
\bar{n}_i \, &=& \, 
\frac{1}{1+[1+(q-1)(\alpha+\beta\,\epsilon_i)]^{\frac{1}{q-1}}}, \cr
\, &=& \, \frac{1}{1+[e_q(-(\alpha+\beta\,\epsilon_i))]^{-1}},
\een

\noindent
where $\alpha$ and $\beta $ are the Lagrange multipliers associated, 
respectively, with the total number of particles and the total energy. 
Notice that we are using here un-normalized $q$-constraints, instead
of using the normalized $q$-constraints studied in \cite{TMP98}. Within the
present context it is not necesary to use the normalized $q$-values,
because {\it our variational problem is formulated in terms of mean 
occupation numbers. These numbers are not probabilities and, consequently, 
do not need to be normalized}. Furthermore, both constraints, the one
associated with the total number of particles, and the one corresponding
to the total energy, are formulated using the same kind of $q$-values. This
allows to re-formulate the variational problem in terms of standard
linear constraints and to prove that the invariance under uniform
shifts of the (single-particle) energy spectrum still holds (see reference
\cite{PPMU04} for details).

Quantum distributions
of the form (\ref{formalsoq}) have been studied by
many researchers in recent years 
\cite{BD93,PPP95,BDG95,TBD98,BD00,WM97,UMK01,TMT03}.
Now, the formal solution
(\ref{formalsoq}) needs to be supplemented with an appropriate
cut-off prescription to deal with negative values of the
quantity $\alpha+\beta\,\epsilon_i$. For $q>1$, our cut-off
prescription is tantamount to adopt in expression (\ref{formalsoq}) a new value 
of $q$\cite{PPMU04},

\begin{equation} \label{qqtwo}
q\longrightarrow q^*=2-q\quad {\rm when}\quad \alpha+\beta\,\epsilon_i \le 0.
\end{equation}

\noindent
 It is interesting to notice that the transformation (\ref{qqtwo})
also arises in other contexts as, for instance, in the study 
of renormalization-group dynamics at the onset of chaos in
logistic maps \cite{BR02a}.

 The most basic requirement of a thermostatitistical
formalism is to be thermodynamically consistent. That is,
the thermostatistical formalism must lead to the standard thermodynamical 
relationships among thermodynamical variables such as entropy, energy, 
temperature, etc \cite{J89,PP93b,TMP98,PP99}.  We are now going to prove that
this is indeed the case with our present formalism.
Let us consider the entropic functional

\begin{equation} \label{qentrop3}
S_q^{(F)}=\sum_i C_q(\bar{n}_i),
\end{equation}

\noindent
where the function $C_q(x)$ is defined by

\begin{equation}
C_q(x)=\left\{\begin{array}{ll}
\left(\frac {x-x^q} {q-1}\right)+\left(\frac{(1-x)-(1-x)^q} {q-1}\right) &\quad 
{\rm if}\quad x\leq \frac{1}{2}\vspace{0.5cm} \\
\left(\frac {x-x^{2-q}} {1-q}\right)+\left(\frac{(1-x)-(1-x)^{2-q}} 
{1-q}\right) &\quad {\rm if}\quad x>\frac{1}{2}
\end{array} \right.
\end{equation}

\noindent
The function $C_q(x)$ is discontinuous at $x=1/2$ (See Figure 2). In fact, the 
limit value of $C_q(x)$ when we approach $x=1/2$ from
the right is

\be
C_q(x\rightarrow 1/2)=\frac{1-2^{1-q}}{q-1},
\ee

\noindent
while the limit value when $C_q(x)$ is approached from the left is,

\begin{equation}
C_q(1/2\leftarrow x)=2^{q-1}\left(\frac{1-2^{1-q}}{q-1}\right).
\end{equation}

\noindent
However, the left and right limit values of the first derivative $dC_q/dx$ at 
$x=1/2$ are both equal to 0. That is, the first derivative $dC_q/dx$ is 
continuous at the cut-off point (See Figure 3).

If we extremalize
the entropic measure (\ref{qentrop3}) under the constraints
(\ref{numeroq}) and (\ref{energiq}) we obtain
the set of equations

\begin{equation}
C_q^{'}(\bar{n}_i)-\alpha\,q\,\bar{n}_i^{q-1}-
\beta\,q\,\epsilon_i\bar{n}_i^{q-1}=0,
\end{equation}

\noindent
which can be solved for the occupation
numbers, yielding

\ben \label{qfernewc}
\bar{n}_i \,
 &=& \, \frac{1}{1+[1+(\tilde 
q-1)(\alpha+\beta\,\epsilon_i)]^{\frac{1}{\tilde q-1}}} \cr
&=& \frac {1} {1+[1+(\tilde q-1)\,\beta(\epsilon_i-\mu)]^{\frac {1} {\tilde 
q-1}}},
\een

\noindent
where $\mu=-\frac{\alpha}{\beta}$ is the chemical potential, and

\be
\tilde q \, = \, \left\{\begin{array}{ll}
q, &\quad {\rm if}\quad \alpha+\beta\,\epsilon_i > 0
\vspace{0.5cm} \\
2-q, &\quad {\rm if}\quad \alpha+\beta\,\epsilon_i \leq 0.
\end{array} \right.
\ee

\noindent
Notice that, 

\begin{equation}
\alpha+\beta\,\epsilon_i=0\quad \Longrightarrow\quad \bar{n}_i=\frac{1} {2}.
\end{equation}

\noindent
The $q$-generalized Fermi-Dirac distribution with our new cut-off prescription 
is depicted, for $q=1,\frac{4}{3}$, and $\frac{5}{3}$ in Figure 4. For 
comparison purposes, the $q$-generalized Fermi-Dirac distribution with the 
standard cut-off rule is exhibited in Figure 5.

When the variational problem is formulated in terms of the entropic measure 
(\ref{qentrop3}), the cut-off prescription need not be imposed on the 
maximum entropy distribution after we obtain the formal
solution of the variational problem. On the contrary,
the solution of the variational problem already contains
the cut-off prescription. We can say that the prescription is incorporated into 
the definition (\ref{qentrop3}) of the entropic functional itself. In terms of 
the generalized exponential function (\ref{exqu}), our $q$-generalized quantum 
distributions for fermions can be written as

\begin{equation} \label{cufer}
\bar{n}_i=\frac {1} {{\rm \tilde e}_q\,[\beta(\epsilon_i-\mu)]\, + \, 1}.
\end{equation}

Now, it can be shown \cite{PP97,Y00} that any thermostatistical formalism
based upon the constrained extremalization of an entropic functional (that is, 
based upon Jaynes maximum entropy approach)
complies with the thermodynamical relationships (which, in the
context of Jaynes' maxent formulation are often referred to as Jaynes' 
relationships). This implies that the $q$-generalization of the Fermi-Dirac
distribution considered in this work, which incorporates the cut-off rule we 
are here advocating, complies with the thermodynamical relationships.
Similar calculations as the ones we have performed here can be done for the 
case of bosons, leading to the $q$-deformed Bose-Einstein distribution,

\begin{equation} \label{cubos}
\bar{n}_i=\frac {1} {{\rm \tilde e}_q\,[\beta(\epsilon_i-\mu)]\, - \, 1}.
\end{equation}

A comment concerning the status of the $q$-deformed quantum distributions 
(\ref{cufer}-\ref{cubos}) is  in order here. This distributions were introduced 
by Buyukkilic and Demirhan (BD)    
\cite{BD93}. BD attempted to derive these distributions in
\cite{BD93} from the full 
$N$-body,( $\Gamma$-space\cite{S})  $q$-generalized
grand canonical ensemble, in a manner similar to  the standard, $q=1$(BG) case \cite{S}.
 It is important to realize that BD's derivation does not 
yield the {\it exact} quantum statistics associated 
with the full $N$-body,( $\Gamma$-space\cite{S})  $q$-generalized
grand canonical ensemble \cite{PPP95}. The BD distributions, 
which have been used by many people \cite{BDG95,TBD98,BD00,WM97,TMT03}, 
can be regarded only as an approximation. Indeed, numerical evidence
has been reported suggesting that, for fermions, the BD distributions
constitute a reasonable approximation in some cases \cite{WM97}.
Furthermore,  a recent BCS s-wave model for high $T_c$
superconductivity, which uses (\ref{cufer}) for the quantum distribution 
functions of the independent quasi-particles \cite{PPMU04}, exhibits 
remarkable agreement with the available experimental data if
a value of the Tsallis parameter $q \sim 1.6$ is adopted \cite{UMK01}. This 
suggests that, at least at some
level,  such quantum distributions might provide a reasonable 
($\mu$-space\cite{S}) description of some quantum many body systems. It is 
important to realize that a similar state of affairs occurs with many of
the other experimental verifications of the $q$-nonextensive
thermostatistics. In most cases a successful account of the experimental data 
has been achieved by recourse to a $\mu$-space theoretical 
model \cite{LK98,B00,AO01,tsallis1}. To determine  the detailed connection 
between these successful $\mu$-space descriptions of concrete physical 
phenomena, on the one hand, and the $\Gamma$-space formulation, on the other, 
still constitutes an open and formidable problem. 
It, therefore, seems prudent, for the time being, to pay serious attention to 
the $\mu$-space  treatments, whenever they reproduce 
experimental data.

\section{Conclusions}

  After considering in detail the usual cut-off prescriptions for the 
$q$-maxent probability distributions, and discussing their most important 
concrete physical realizations, the following conclusions are inescapable. 
First of all, the main physical realizations of the standard cut-off rule 
correspond to
the case $q<1$. Secondly, the usual prescription
for $q>1$ is physically not as interesting classically as
the usual cut-off rule for $q<1$. In the present
effort we proposed an alternative rule for the case $q>1$. We illustrated our 
proposal with $q$-generalized quantum distributions that have already been 
successfully applied
to the study of both high $T_c$ superconductivity \cite{UMK01} and the 
formation of the quark-gluon plasma \cite{TMT03}. In the
particular example of the $q$-generalized quantum distributions we have proved 
that our
cut-off prescription leads to a thermodynamically consistent formalism.
 Even though we have focused here upon Tsallis' generalized thermostatistics, 
it has not escaped our attention that the present considerations may be also 
relevant for other non-standard thermostatistical formalisms that have recently
been proposed \cite{LMS00,K02}.

Further developments and applications of our variable-$q$ Tsallis formalism, 
and the exploration of its relationship with problems, will be greatly welcome.

\acknowledgements
The support of the Foundation for Research Development of South Africa is 
gratefully acknowledged. 


\vskip 0.5cm

\newpage

\begin{figure}
\centering
\begin{center}
\includegraphics[scale=0.4,angle=270]{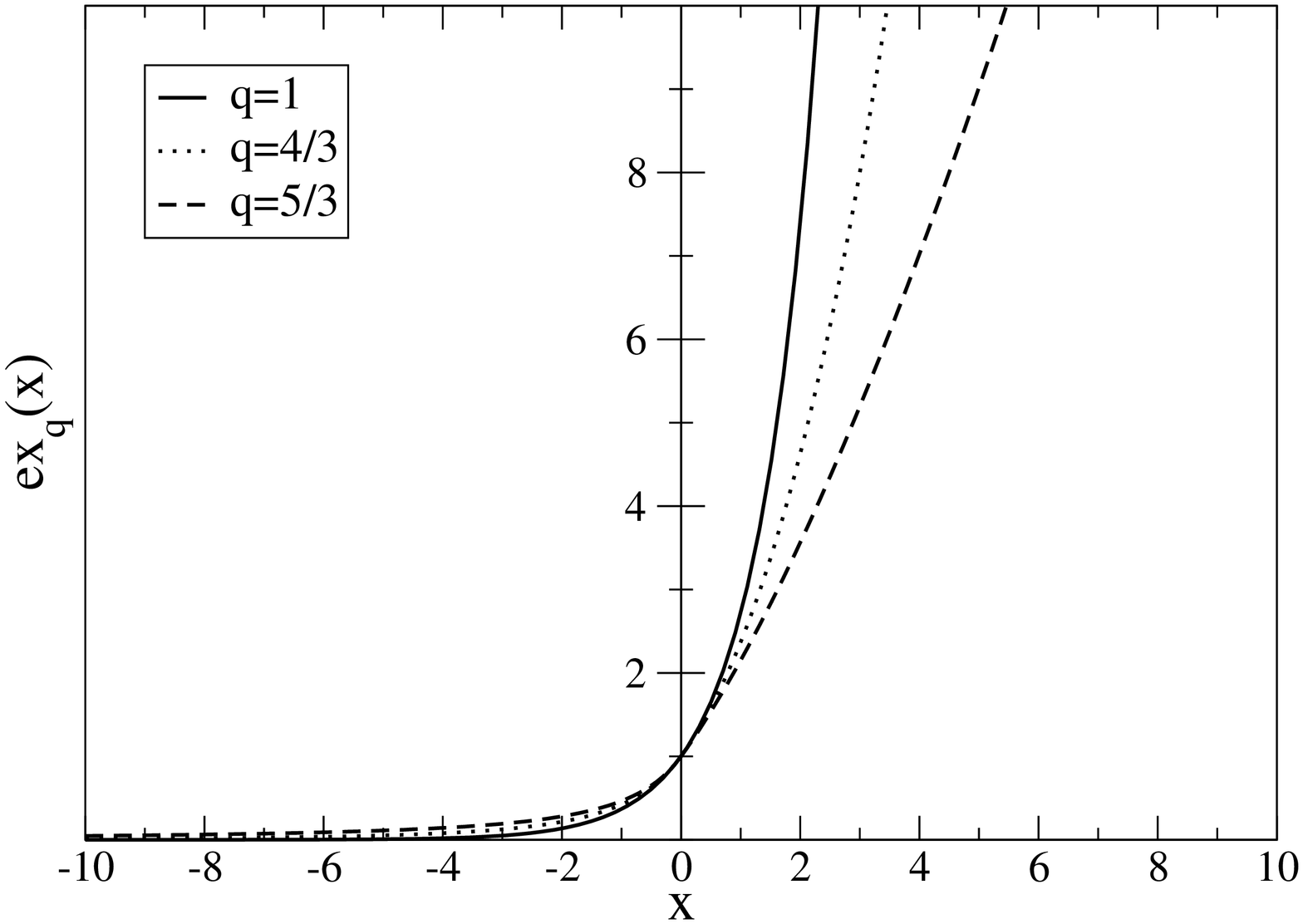}
\end{center}
\label{fig1}\caption{ The generalized exponential function
$ex_q(x)$ for $q=1,4/3,5/3$. All depicted quantities are dimensionless.
 }
\end{figure}
\begin{figure}
\centering
\begin{center}
\includegraphics[scale=0.4,angle=270]{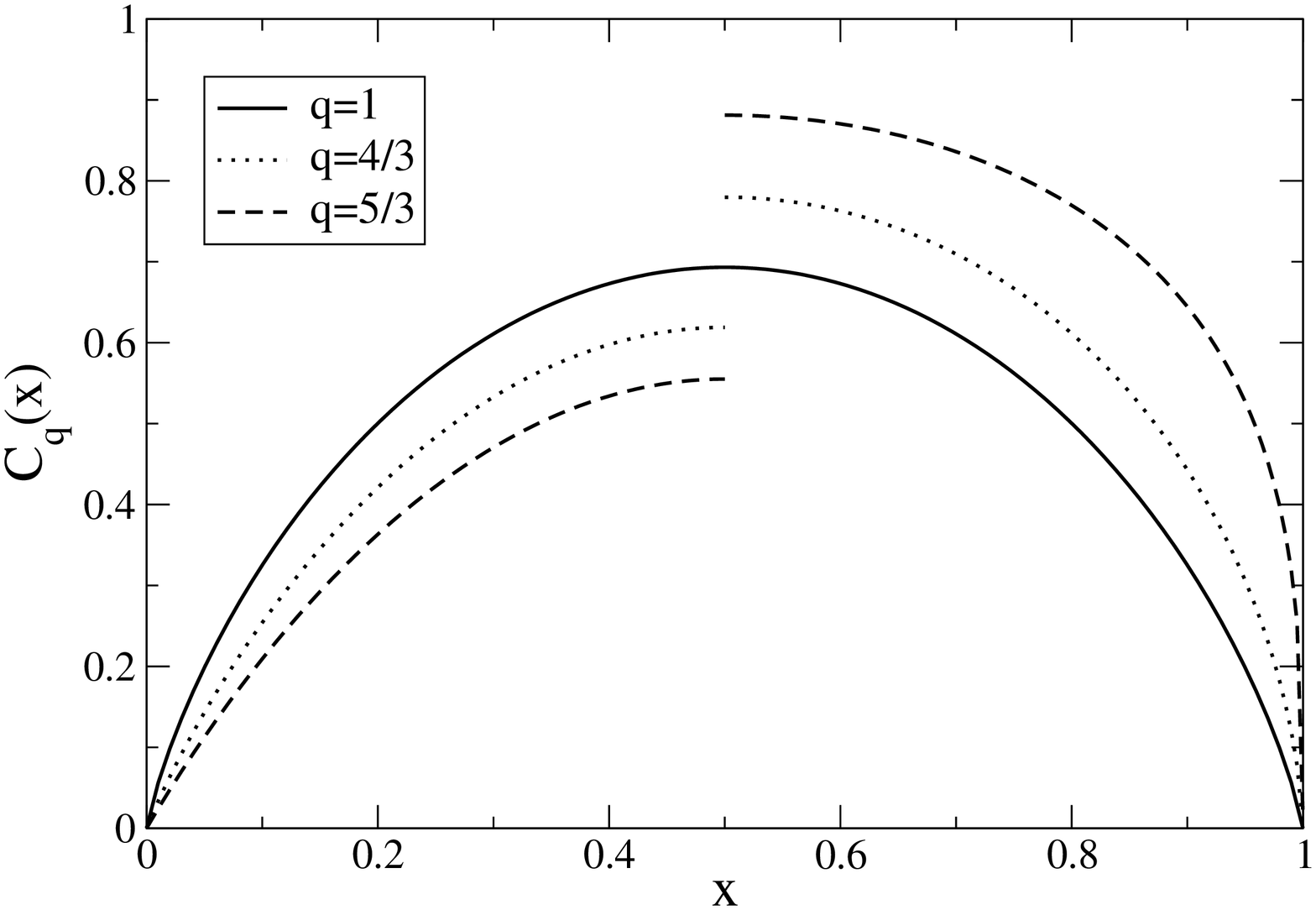}
\end{center}
\label{fig1}\caption {The function $C_q(x)$, appearing in
the definition of the entropic functional, for $q=1,4/3$, and $5/3$.
All depicted quantities are dimensionless.
}
\end{figure}
\begin{figure}
\centering
\begin{center}
\includegraphics[scale=0.4,angle=270]{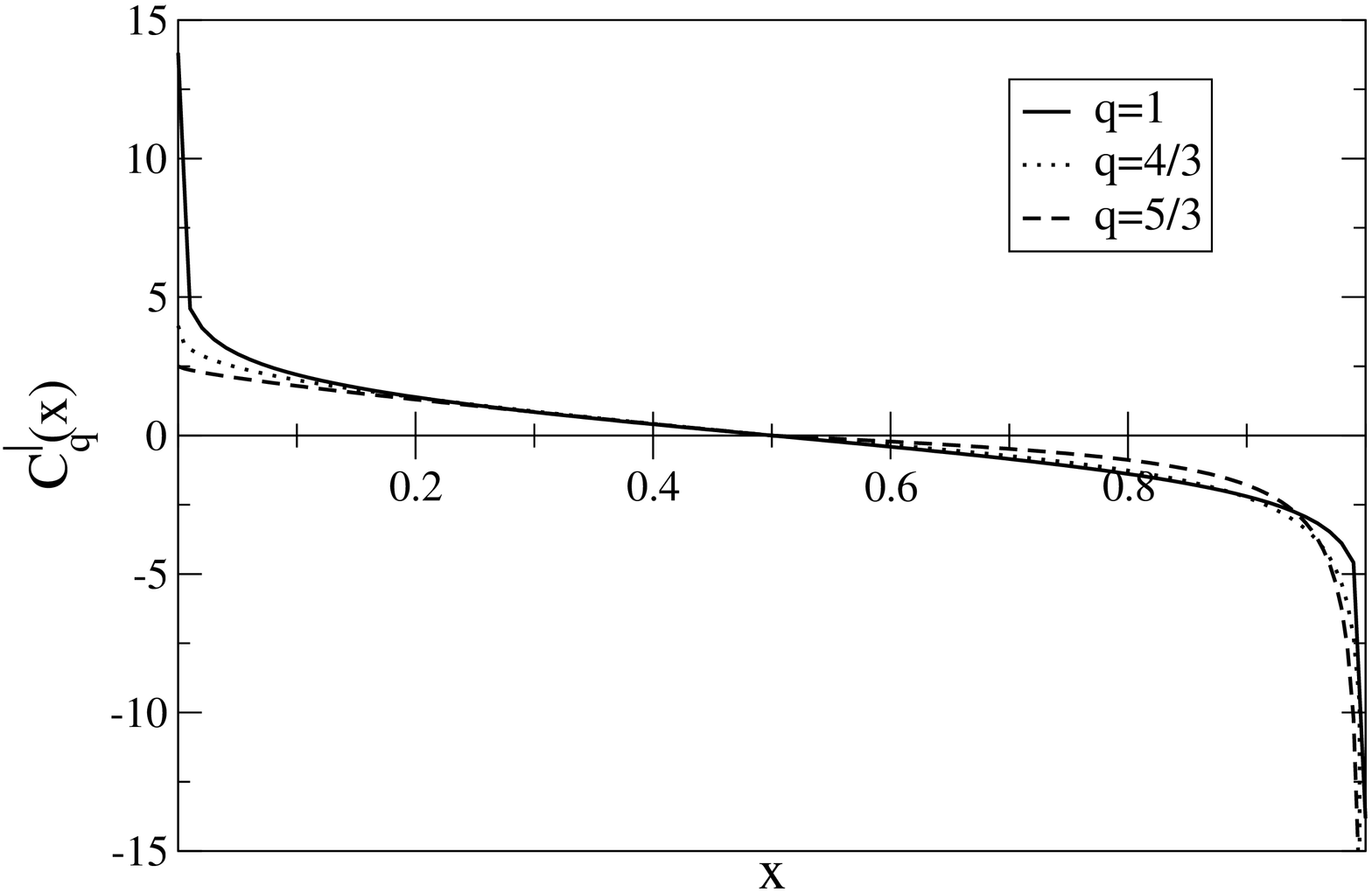}
\end{center}
\label{fig1}\caption {The derivative $dC_q(x)/dx$ of the function $C_q(x)$ appearing 
in
the definition of the entropic functional, for $q=1,4/3$, and $5/3$.
All depicted quantities are dimensionless.
}
\end{figure}
\begin{figure}
\centering
\begin{center}
\includegraphics[scale=0.4,angle=270]{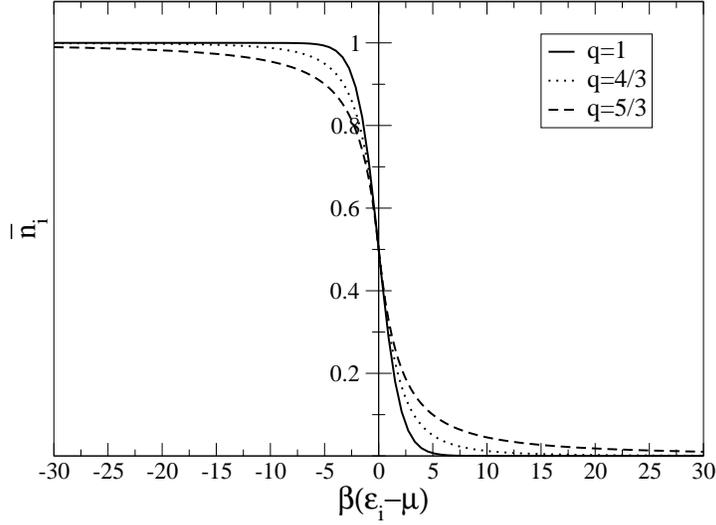}
\end{center}
\label{fig1}\caption {The average ocupation number $n_i$
for the $q$-generalized Fermi-Dirac distribution
(\ref{qfernewc}), which incorporates our new cut-off prescription,
as a function of $\beta(\epsilon_i-\mu)$ and for
the same $q$-values as in Figures (1-3).
 All depicted quantities are dimensionless.
}
\end{figure}
\begin{figure}
\centering
\begin{center}
\includegraphics[scale=0.4,angle=270]{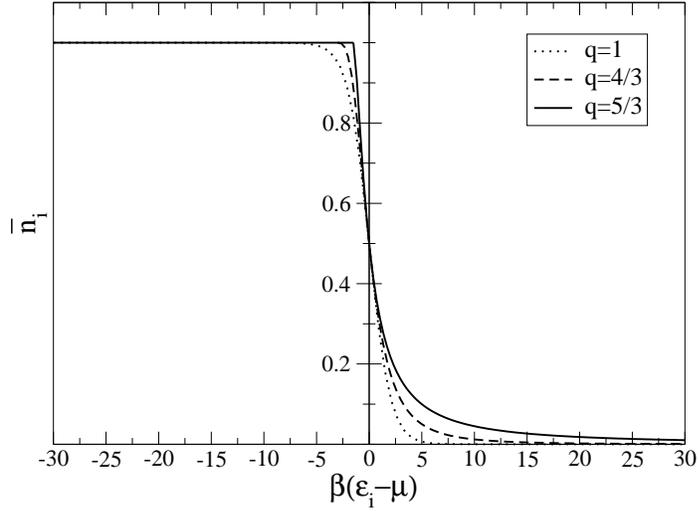}
\end{center}
\label{fig1}\caption {The average ocupation number for the $q$-generalized 
Fermi-Dirac distribution
with the standard cut-off prescription, as a function of 
$\beta(\epsilon_i-\mu)$ and for
the same $q$-values as in Figures (1-3).
 All depicted quantities are dimensionless.
}
\end{figure}

\end{document}